\documentclass{aa}  
\usepackage{graphicx}
\usepackage{txfonts}
\usepackage{natbib}
\usepackage{dcolumn}
\newcolumntype{.}{D{.}{.}{-1}}
\newcolumntype{;}{D{;}{.}{7}}
\bibpunct{(}{)}{;}{a}{}{,} % to follow the A&A style

\newcommand{\solm}{M$_{\odot}$\ }

\begin{document}

\authorrunning{Perger et al.}
\titlerunning{Compact GC MIR sources}
\title{Compact mid-IR sources east of galactic center source IRS5}
\subtitle{}
\author{M. Perger\inst{1}, J. Moultaka\inst{1,2}, A. Eckart\inst{1,3}, T. Viehmann\inst{1}, R. Sch\"odel\inst{1}
          \and
K. Muzic\inst{1,3}
          }
\offprints{A. Eckart (eckart@ph1.uni-koeln.de)}

\institute{I. Physikalisches Institut, Universit\"at zu K\"oln, Z\"ulpicher Str. 77, 50937 K\"oln, Germany
         \and
                Laboratoire d'Astrophysique de Toulouse, UMR 5572, Observatoire
               Midi-Pyr\'en\'ees, 14 avenue Edouard Belin, 31400
               Toulouse, France
         \and
                Max-Planck-Institut f\"ur Radioastronomie, Auf dem H\"ugel 69, 53121 Bonn, Germany
             }

%\institute{

\date{Received 9. Aug. 2007 / Accepted 25. Oct. 2007}

  \abstract
{}
{Mid-infrared observations of the Galactic Center show among the extended mini-spiral 
a number of compact sources. Their nature is of interest because they represent an interaction 
of luminous stars with the mini-spiral material or mass losing sources that are enshrouded in dust and gas shells. 
Characterizing their nature is necessary in order to obtain a complete picture of the different stellar populations 
and the star formation history of the cental stellar cluster in general.
Prominent compact MIR sources in the Galactic Center are either clearly offset from the mini-spiral (e.g. the M2 super-giant IRS~7 
and the bright dust enshrouded IRS~3) or have been identified earlier with bright bow shock sources (e.g.
IRS~21, 1W, 10W and IRS~5).
There are, however, four less prominent compact sources east of IRS~5, the natures of which were unclear until now.
}
  % methods heading (mandatory)
{We present near-infrared K-band long slit spectroscopy of the four sources east of IRS~5
obtained with the 
ISAAC spectrograph
at the ESO VLT in July 2005.
We interpret the data in combination with high angular resolution NIR and MIR images 
obtained with ISAAC and NACO at the ESO VLT.}
  % results heading (mandatory)
{The K'-band images and proper motions show that the sources are multiple. 
For all but one source we find dominant contributions from late type stars with best 
overall fits to template stars with temperatures below 5000~K.}
  % conclusions heading (mandatory)
{The brightest sources contained in IRS~5NE, 5E and 5S may be  
asymptotic giant branch stars and a part of the MIR
excess may be due to dust shells produced by the individual sources. 
However, in all cases an interaction with the mini-spiral cannot be excluded and their broad band 
infrared SEDs indicate that they could be 
lower luminosity counterparts of the identified bow shock sources.
In fact, IRS~5SE is associated with a faint bow shock and its spectrum shows 
contributions from a hotter early type star which supports such a classification.}


\maketitle
\maketitle
%
%________________________________________________________________

\section{Introduction}
The Galactic Center (GC) 
at a distance of $\sim$8~kpc (Ghez et al. 2005;
Sch\"odel et al. 2002, 2003;
Eisenhauer 2003, 2005)
is known as a bright source of near- and mid-infrared (NIR and MIR) radiation since the late 
1960s (\cite{Becklin1}; \cite{Becklin2}; \cite{Low}). 
The main source of NIR radiation is photospheric emission from a dense stellar cluster, i.e. a crowded field of point sources, while almost all of the MIR radiation originates from extended gas and dust features as well as dust emission from the circum-stellar regions of a dozen individual sources interacting with the more extended GC interstellar medium (ISM). 

\begin{figure*}
   \centering
   \includegraphics[width=18cm]{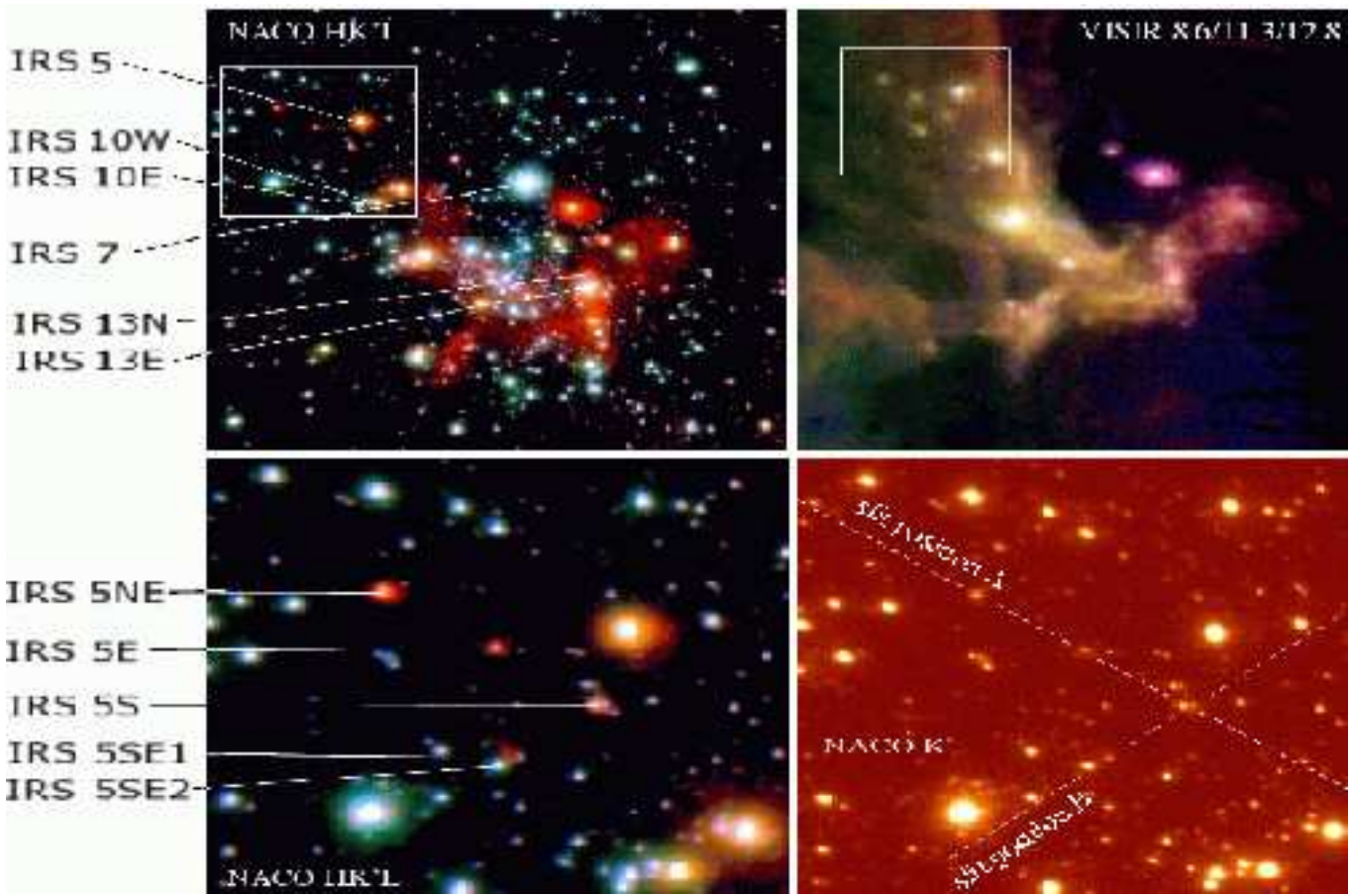}
      \caption{Upper left: NACO HK'L-image of the inner 30$''\times$30$''$ region of the GC. 
Upper right: VISIR N-band three-color composite view of the GC. 
The blue channel is 8.6~$\mu$m, green is 11.3~$\mu$m, and red is 12.8~$\mu$m (30$''\times$ 30$''$). 
Lower left: NACO HK'L-image of the observed sources (6$''\times$6$''$). 
A more detailed nomenclature of sources is given in 
Table~\ref{pmtab} and Fig.~\ref{figpm1}.
Lower right: NACO K'-band image of the observed sources (6$''\times$6$''$).
The white insets in the top panels indicate the location of the enlarged area shown in the 
lower panels. The dashed white lines in the lower right panel indicate the orientation of the
0.6$''$ ISAAC slits.}
         \label{FigVibStab8}
   \end{figure*}

The GC stellar cluster shows some intriguing characteristics: 
it is extremely dense, with an unusual observed stellar population (\cite{Genzel03}; \cite{Eisenhauer05}).
Recently, Sch\"odel et al. (2007) presented AO assisted high-resolution NIR imaging observations of the stellar 
cluster within 20$''$ (about 0.75~pc) of Sgr A*, the massive black hole at the center of the Milky Way.
Sch\"odel et al. (2007) extracted stellar number counts and colors, and derived from them the detailed structure 
of the nuclear stellar cluster and an extinction map across it.
The bright members of the central stellar cluster are mainly 
(80\% of all m$_{K} \leq$ 14 mag stars; \cite{Ott}) late-type red giants (e.g. IRS~7 and IRS~10E in Fig.~\ref{FigVibStab8}), many of which are suspected
to lie on the asymptotic giant branch (AGB). It is also composed of young massive stars which have 
energetic winds (e.g. \cite{Krabbe}; \cite{Najarro}; IRS~13E in Fig.~\ref{FigVibStab8})
and are arranged in two stellar disks (\cite{Genzel03}; \cite{Paumard06}). 
Spectra of AGB stars show strong 2.3~$\mu$m CO bandhead absorption and the massive, 
hot and windy stars (He-stars) exhibit He/H emission. These emission line stars dominate the NIR luminosity of the central few arcseconds. 
All bright and compact MIR sources in the GC are either clearly offset from the mini-spiral (such as the M2 super-giant IRS~7
and the bright dust enshrouded IRS~3) or have been identified earlier with bright bow shock sources (like
IRS~21, 1W, 10W and IRS~5).

 \begin{figure}
  \centering
  \includegraphics[width=8cm]{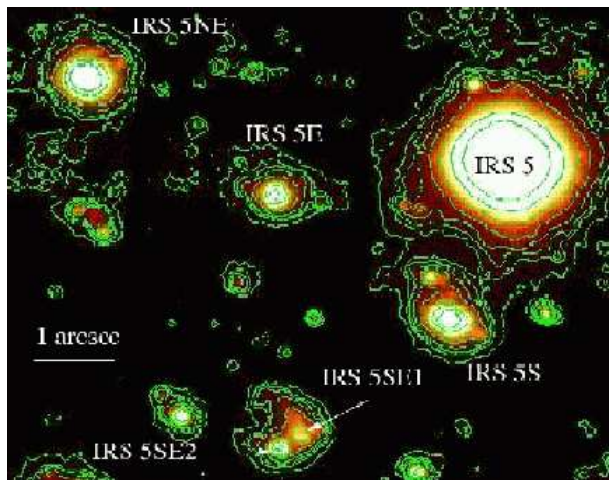}
  \caption{An L'-band shift-and-add adaptive optics 
image showing the compact MIR sources east of
IRS~5 with logarithmic contours. The extended bow shock like structure 
to the northeast  of IRS~5SE1 is clearly visible.}
  \label{Figzoom}
  \end{figure}

A third, less numerous component of the GC stellar cluster, consists of luminous objects with steep, red and 
featureless (K-band-) spectra and a strong infrared excess.
Although clearly extended in MIR images they are quite compact in this wavelength range compared to 
the mini-spiral or the dust shell around IRS~3 (\cite{Viehmann06}).
These clearly extended dust embedded MIR sources (e.g. IRS~5, 10W in Fig.~\ref{FigVibStab8}) are bow shock sources, caused by bright emission-line stars with strong winds plowing through the ambient gas and dust of the northern arm of the mini-spiral (\cite{Tanner1}; \cite{Tanner2}; \cite{Tanner3}; \cite{Rigaut}; \cite{Geballe1}; \cite{Geballe2}). 
There are also fainter dust embedded sources 
(e.g. the IRS~13N cluster in Fig.~\ref{FigVibStab8}; \cite{Eckart}) which could represent a
low luminosity class of bow shock sources or even young dust embedded stars that have been recently formed. 
In this paper we investigate compact MIR sources close to or within the northern arm of the mini-spiral.

Infrared images longward of 2~$\mu$m show four bright compact objects (\cite{Viehmann06}) 
located to the east of the bright northern arm source IRS~5. They are especially prominent in
high angular resolution MIR images as obtained with VISIR (see Fig.~\ref{FigVibStab8}
and a zoom towards the described sources in Fig.~\ref{Figzoom}). 
These sources have an exceptional appearance  since they are quite compact compared to the 
bulk of the 10~$\mu$m emission which is associated with the extended mini-spiral. 
The nature of these four compact sources, which we refer to as IRS~5NE, 5E, 5S and 5SE 
(see Fig.~\ref{FigVibStab8}; IRS~5SE includes 5SE1 and 5SE2), is currently unclear: they are almost as bright as IRS~7 
(M2 supergiant in Fig.~\ref{FigVibStab8}; \cite{Carr}) in the N-band, while they appear 
much less prominent (but still detectable) at shorter wavelengths (K-band), 
although they remain bright in L- and M-bands. 
%A reliable classification and/or understanding of these sources can only be achieved
%through spectroscopic methods supported by high angular resolution imaging. 
We present first high angular resolution images, proper motions and K-band long-slit spectroscopy as obtained with the ESO VLT, 
in order to further investigate the properties of these sources.

\section{Observations and data reduction}

\subsection{Proper motion data}
\begin{figure}
   \centering
   \includegraphics[width=09.1cm]{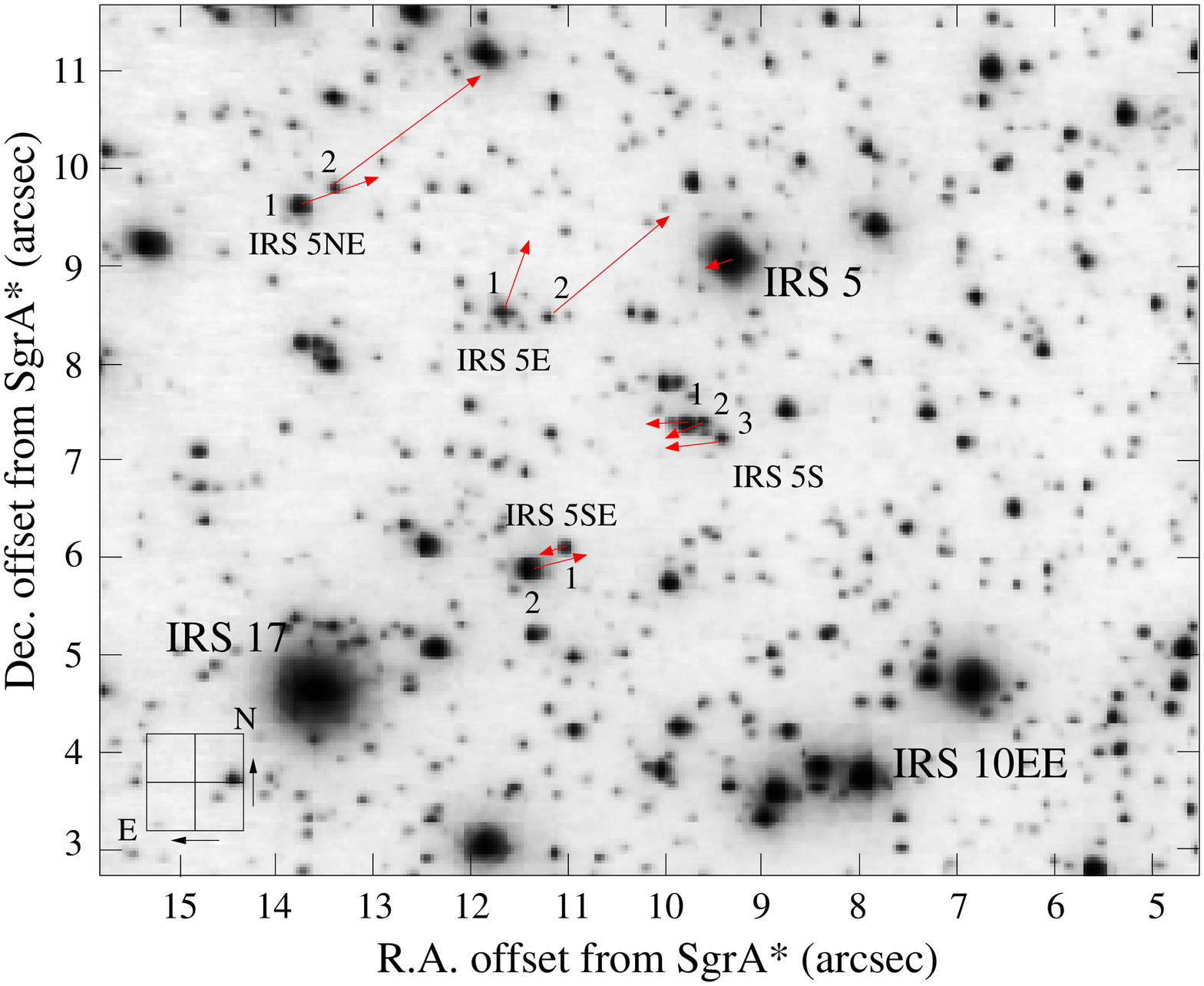}
      \caption{Proper motions of the compact MIR excess sources 
east of the northern arm of the mini-spiral.} 
         \label{figpm1}
   \includegraphics[width=09.1cm]{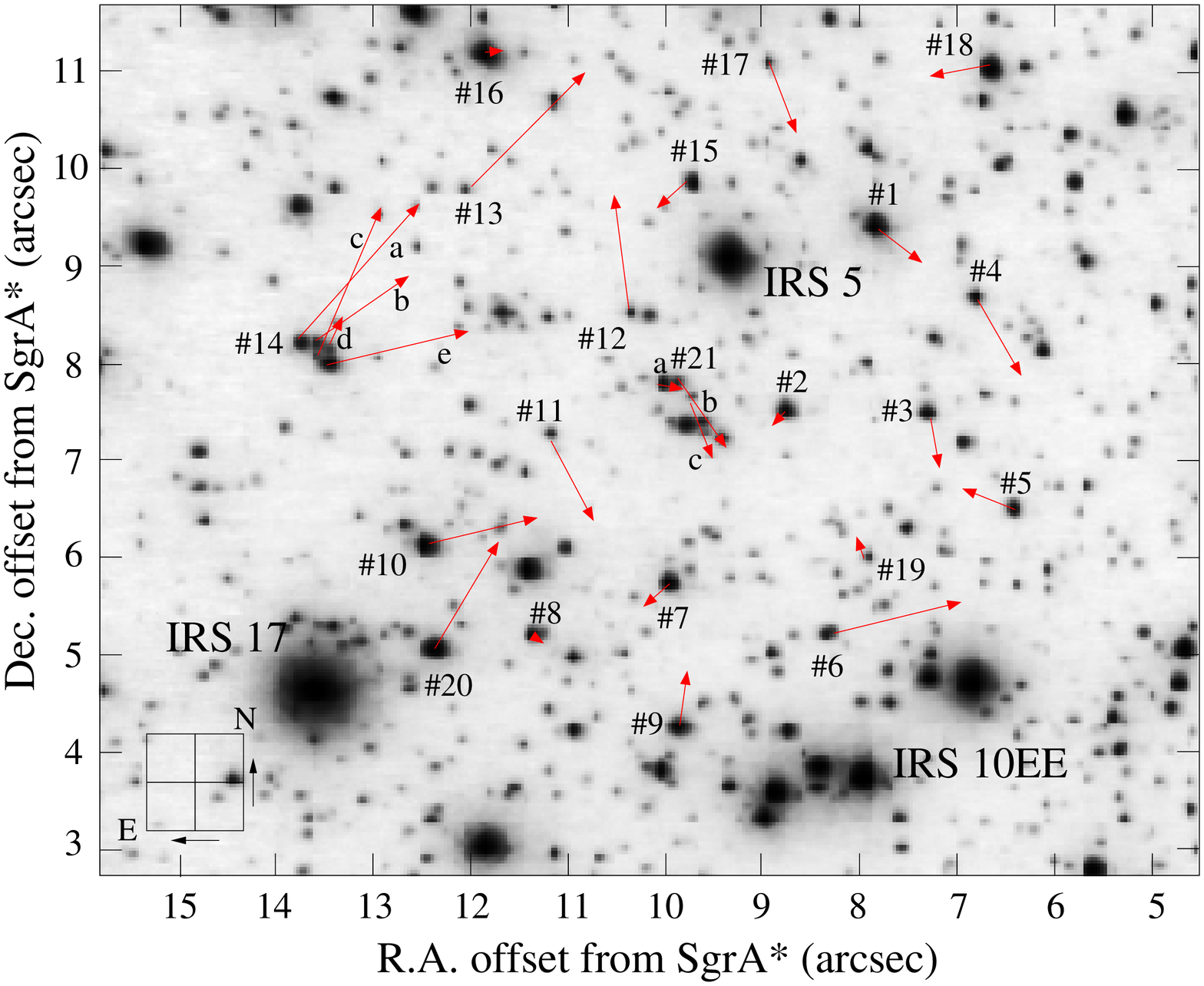}
      \caption{Proper motions of randomly selected sources to the east of IRS~5.
In both panels (Figs.~\ref{figpm1} and \ref{figpm2}) the square in the lower left
corner indicates the orientation of positive velocity vectors and has a
total width of 200~km/s in both directions.}
         \label{figpm2}
   \end{figure}

The K'-band (2.18~$\mu$m) images from which we derived the proper motions of the sources 
were taken with the NAOS/CONICA\footnote{Nasmyth Adaptive Optics System/COude Near Infrared CAmera; \cite{lenzen03}; \cite{rousset03}} adaptive optics assisted imager/spectrometer 
(\cite{lenzen98}; \cite{rousset98}; \cite{brandner02}
at the UT4 (YEPUN) at the ESO\footnote{European Southern Observatory} VLT\footnote{Very Large Telescope}, Cerro Paranal, Chile.
The data set includes the images from epochs 2002.339, 2003.356, 2004.512, 2004.521 and 
2006.413\footnote{ESO program IDs: 60.A-9026(A), 073.B-0775(A,B), 077.B-0552(A)} with an angular resolution of $\sim$56~mas.
Data reduction (bad pixel correction, sky subtraction, flat field correction) 
and formation of final mosaics was performed using the DPUSER software for 
astronomical image analysis (T. Ott; see also \cite{eckart90}).  
The absolute positions of sources in our AO images were derived by comparison to the
VLA positions of IRS~10EE, 28, 9, 12N, 17, 7 and 15NE as given by \cite{menten97} and \cite{reid03} (for identification of different GC sources see e.g. Fig.~2 in \cite{Viehmann05}).
The radio positions and the positions in the K'-band image agree to within 
less than a single 27~mas pixel i.e. less than a K'-band diffraction limited beam.
The stars used for
the transformation were chosen to be uniformly distributed across the field. 
In addition to the MIR excess sources east of IRS~5 we have randomly selected 27 objects in the field
for comparison.
The positions and K'-band proper motions of all sources are listed in Table~\ref{pmtab}
and shown in Figs.~\ref{figpm1} and \ref{figpm2}. 
The proper motions of the overall source sample in Table~\ref{pmtab} are in agreement with random motions.
Within the width of the velocity distributions of about $\sigma_{field}$=150 km/s 
their median velocities of about -63 km/s to the east and 2 km/s to the north are 
close to zero with random orientations.

\begin{table}[htb]
\caption{Proper motion data (and their 1$\sigma$ uncertainties) of the compact IRS~5 N-band sources
derived from NACO K'-band data,
covering the epochs 2002.339, 2003.356, 2004.512, 2004.521 and 
2006.413 with a resolution of $\sim$56~mas. Positive velocities go
from west to east and south to north. 
The epoch 2002 K'-band positions are referred to SgrA* with a 1$\sigma$ uncertainty of 0.1$''$ }             
\begin{center}
\begin{tabular}{lrrcc}\hline \hline
Source &$\Delta_{\alpha}$ &$\Delta_{\delta}$& v$_{R.A.}$  & v$_{Dec}$ \\  
       &        [$''$]                &           [$''$]            & [km/s] & [km/s] \\ \hline 
IRS~5   &  9.31& 9.15 &   43 $\pm$   18& -27 $\pm$   12\\ 
IRS~5NE1& 13.79& 9.60 & -156 $\pm$   34&  82 $\pm$   20\\ 
IRS~5NE2& 13.31& 9.88 & -346 $\pm$   10& 279 $\pm$   12\\ 
IRS~5E1 & 11.68& 8.50 &  -48 $\pm$   30& 183 $\pm$   25\\ 
IRS~5E2 & 13.31& 9.88 & -299 $\pm$   10& 216 $\pm$   12\\
IRS~5S1 &  9.80& 7.31 &  105 $\pm$   14&   2 $\pm$   13\\ 
IRS~5S2 &  9.62& 7.32 &  102 $\pm$   14& -34 $\pm$   16\\ 
IRS~5S3 &  9.42& 7.19 &  179 $\pm$   22& -46 $\pm$   16\\ 
IRS~5SE1& 10.00& 6.06 &   60 $\pm$   19& -22 $\pm$   16\\ 
IRS~5SE2& 11.30& 5.91 & -150 $\pm$   10&  24 $\pm$   10\\ \hline
$\#$~1  &  7.82& 9.42 & -113 $\pm$   10& -69 $\pm$   11\\
$\#$~2  &  8.75& 7.52 &   28 $\pm$    9& -26 $\pm$   10\\
$\#$~3  &  7.22& 7.50 &  -20 $\pm$   12& -96 $\pm$   12\\
$\#$~4  &  6.87& 8.70 & -102 $\pm$   11&-181 $\pm$   11\\
$\#$~5  &  6.45& 6.50 & -112 $\pm$   15&  45 $\pm$   16\\
$\#$~6  &  8.25& 5.20 & -299 $\pm$   52&  79 $\pm$   25\\
$\#$~7  &  9.98& 5.80 &   69 $\pm$    9& -46 $\pm$   10\\
$\#$~8  & 11.27& 5.21 &    9 $\pm$   22&  -5 $\pm$   16\\
$\#$~9  &  9.90& 4.20 &  -17 $\pm$    9& 106 $\pm$   11\\
$\#$~10 & 12.47& 6.11 & -247 $\pm$    8&  48 $\pm$   10\\
$\#$~11 & 11.18& 7.19 &  -96 $\pm$   10&-170 $\pm$   12\\
$\#$~12 & 10.30& 8.62 &   18 $\pm$    9& 228 $\pm$   10\\
$\#$~13 & 12.05& 9.87 & -263 $\pm$   10& 222 $\pm$   12\\
$\#$~14a& 13.80& 8.22 & -308 $\pm$   16& 322 $\pm$   15\\
$\#$~14b& 13.67& 8.23 & -210 $\pm$   20& 127 $\pm$   18\\
$\#$~14c& 13.67& 8.08 & -139 $\pm$   32& 304 $\pm$   24\\
$\#$~14d& 13.46& 8.12 &  -46 $\pm$   49&  64 $\pm$   32\\
$\#$~14e& 13.40& 8.00 & -334 $\pm$   10&  65 $\pm$   11\\
$\#$~15 &  9.79& 9.92 &   79 $\pm$    9& -50 $\pm$   10\\
$\#$~16 & 11.93&11.20 &  -50 $\pm$    8&   3 $\pm$   10\\
$\#$~17 &  8.93&11.10 &  -63 $\pm$   14&-161 $\pm$   15\\
$\#$~18 &  6.62&11.05 &  109 $\pm$   11&  -9 $\pm$   11\\
$\#$~19 &  7.93& 6.00 &   14 $\pm$   10&  45 $\pm$   11\\
$\#$~20 & 12.32& 5.04 & -165 $\pm$   10& 218 $\pm$   11\\
$\#$~21a& 10.00& 7.83 &  -69 $\pm$   13& -14 $\pm$   13\\
$\#$~21b&  9.87& 7.82 & -102 $\pm$   23&-157 $\pm$   17\\
$\#$~21c&  9.75& 7.62 &  -53 $\pm$   12&-111 $\pm$   12\\
\hline \hline
\end{tabular}
\end{center}
\label{pmtab}
\end{table}

\subsection{ISAAC observations}
On July 27$^{th}$ and 28$^{th}$ 2005\footnote{ESO program ID: 075.C-0138(A)}, K-band (2.2~$\mu$m) spectroscopy observations of the 
four MIR sources east of IRS~5 (m$_{K,obs} \sim$ 8 to 11 mag)
were obtained with the ISAAC\footnote{Infrared Spectrometer And Array Camera, \cite{Moorwood}} 
infrared spectrometer mounted at the UT1 (ANTU) at the 
ESO VLT. The seeing-limited spectra (0.7 to 1.6$''$) 
were taken with a long-slit of 0.6 x 120$''$ (two slit positions A and B, see Fig.~\ref{FigVibStab8}) 
in ISAAC$`$s lower resolution mode (R=$\lambda$/$\Delta\lambda$=840). 
The observations were made with an integration time of 100~sec per frame with total integration times
of 2400~sec for IRS~5S and 1200~sec for the remaining 3 objects IRS~5NE, 5SE, and 5E. 

The spectroscopy data reduction was carried out with IRAF\footnote{Image Reduction and Analysis Facility, 
distributed by the National Optical Astronomy Observatory (NOAO), operated by the 
Association of Universities for Research in Astronomy, Inc. (AURAI) and under cooperative 
agreement with the National Science Foundation (NSF)} \& IDL\footnote{Interactive Data Language} 
using standard procedures: the sky-subtraction was achieved with a nodding technique and 
flat-fielding was applied. The individual spectra were corrected for slit-curvature and shifts. 
Creating a median of individual spectra then resulted in the final spectrum for a given source. 
Telluric correction was performed with A- and B-type stars, the observations of which 
directly followed or preceded those of the targets. The hydrogen absorption line 
(Br$\gamma$ at 2.17~$\mu$m) in the stellar template spectra was removed by fitting a 
Lorentz-profile to the line prior to the telluric correction. 
At wavelengths of less than 2~$\mu$m we used the underlying black body 
continuum of the star. Consequently no line correction was applied.
After division by the 
spectrum of the telluric standards, the spectra of the four MIR sources east of IRS~5 were 
multiplied by a blackbody of temperature equal to the effective temperature of the standard stars. 
The spectra were wavelength calibrated using Xenon-lines from additionally observed data frames. 
Flux density calibration was applied with zero points from literature 
(FD$_{0,K}$ = 657 Jy, \cite{Skinner}) and high resolution NACO images (compared to IRS~16NE and 16NW). 
Strong atmospheric CO$_{2}$ absorption 
result in lower signal-to-noise ratios (SNRs) in the
1.98 to 2.06~$\mu$m spectral regions.
At shorter wavelengths the SNR of the spectra suffer from 
residual telluric H$_2$O absorption and the 
strong ($\sim$30~mag; see below) extinction towards the GC region.

\section{Results}

\subsection{Structures and proper motions of individual sources}

In this section we discuss 
the structures and proper motions of the individual 
compact MIR sources east of IRS~5 as obtained from
our high angular resolution K'-images.
\\
{\bf IRS~5NE:} The K'-band images show two sources 
with a separation of 0.4$''$. 
Both sources are moving towards the northwest but their velocity difference
is large with respect to the uncertainties and $\sigma_{field}$.
This implies that the two sources are not physically associated.
\\
{\bf IRS~5E:} The situation is similar to that of IRS~5NE. 
IRS~5E1 and IRS~5E2 are an apparent double source with an east-west 
separation of 0.5$''$ and a large proper motion velocity difference. 
IRS~5E2 is about half as bright as IRS~5E1.
Both sources are moving towards the northwest. 
The images also indicate several sources 
which are at least 1 magnitude fainter than 
IRS~5E1 and have separations of less 
than 0.5$''$ from it. 
They are too faint to determine their proper motions.
\\
{\bf IRS~5S:}
Here a minimum of 3 sources within a 0.6$''$ diameter region is moving towards the east-southeast.
Their proper motion velocity difference
is small with respect to the uncertainties and $\sigma_{field}$.
This implies that these sources could be physically associated.
A firm statement on this requires a determination of their radial velocities in the near future.
\\
{\bf IRS~5SE:}
High-resolution NAOS/CONICA K'- and L-band images (see Fig.~\ref{FigVibStab8}) reveal that the
southernmost of the four sources, which appears slightly extended 
($\approx$0.2$''$; compared to IRS7) in the VISIR images,
is in fact an apparent double source, consisting of a 'blue' point source to the east (SE2) 
and a fainter point source to the west (SE1). 
The spectrum we obtained from IRS~5SE probably includes flux density contributions from
both objects.
We calculated a separation between the two objects as 0.4$\pm$0.1$''$.
Their proper motion  velocity difference
is large with respect to the uncertainties and $\sigma_{field}$.
This implies that the two sources are not physically associated.
IRS5~SE1 shows a tail like structure (in the L'-band)
that appears to be the main source of emission at longer wavelengths. 
The extended dust feature has the appearance of a bow-shock.
Since IRS5~SE2 is blue and bright it may be associated with a luminous star
that interacts with the surrounding ISM. 
The position angle of the bow shock with respect to the proper motion of the
source it is physically associated with, depends on the geometry and density 
structure of the local ISM it is moving in. Here offsets by $\sim$45$^o$ 
appear to be possible (see Fig.~\ref{Figzoom}).

\subsection{Stellar types}
\label{Stellartypes}

  \begin{figure}
   \centering
   \includegraphics[height=6cm, width=9cm]{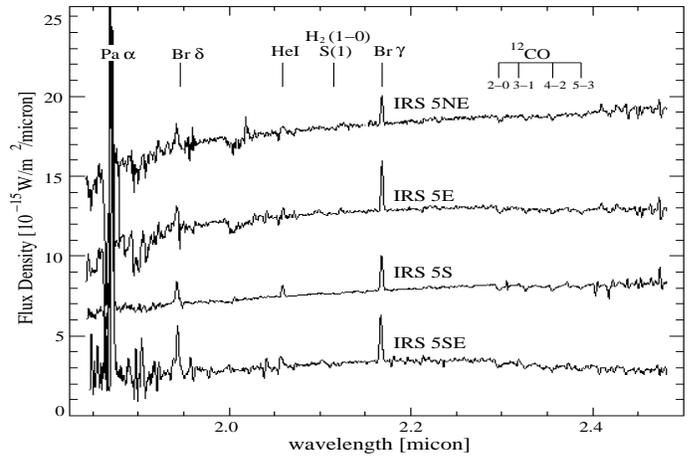}
      \caption{The fully reduced spectra of IRS~5NE, 5E, 5S and 5SE. The most prominent absorption and emission lines are marked. The flux densities are shifted for display purposes (IRS~5NE: +15$\times 10^{-15} W /m^{2} /  \mu m$, 5E: +10$\times 10^{-15} W /m^{2} / \mu m$, 5S: +5$\times 10^{-15}W /m^{2} / \mu m$, 5SE: -5$\times 10^{-15} W /m^{2} / \mu m$). 
Note the lower SNR for wavelengths smaller than 1.94~$\mu$m and at around 2.02~$\mu$m.
For spectral flux calibration - see~2.2.
}
         \label{FigVibStab}
   \end{figure}

  \begin{figure}
   \centering
   \includegraphics[height=15cm,width=8cm]{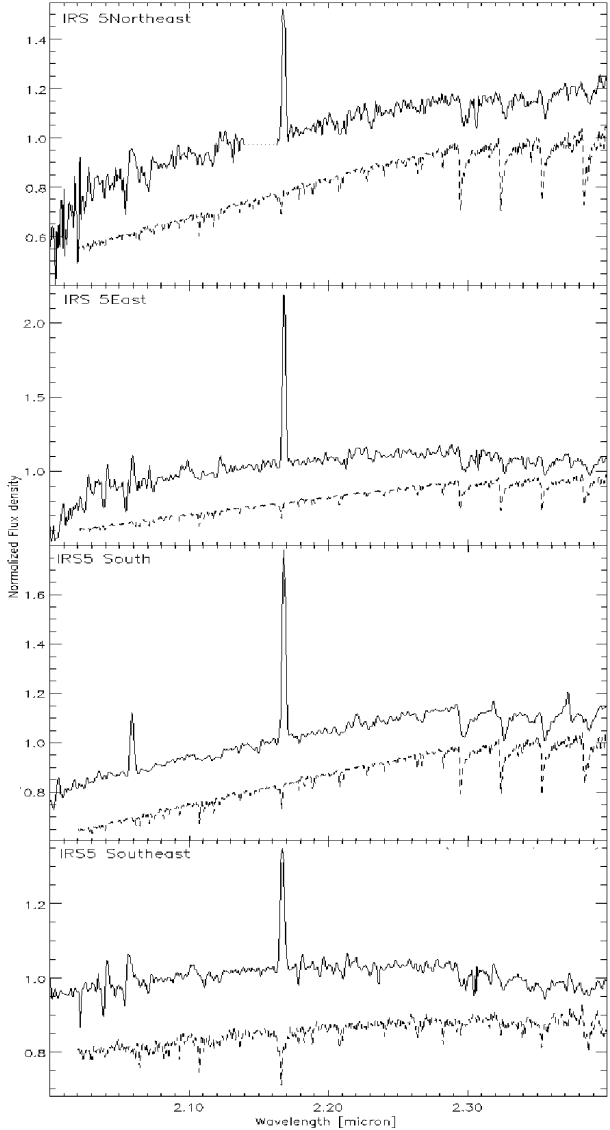}
      \caption{Flux density normalized reduced spectra (upper curve) and flux density normalized and extincted 
standard star spectra (lower curve) by Wallace \& Hinkle (1997) which show the smallest 
differences in magnitude and in the CO bandhead depth and EW. 
The latter are shifted along the flux density axis for display purposes.
For IRS~5SE the identification of a strong early type stellar contribution to the 
spectrum is also consistent with a reduced Br$\gamma$ emission strength at the wavelength of the
corresponding absorption in the template spectrum.}
         \label{FigVibStab2}
   \end{figure}

\begin{table*}
\caption{Extinction corrected flux densities (FD) of the compact N-band sources in Janskys 
(\cite{Viehmann06}) and H-, K- and L-band flux densities and magnitudes (m) from NACO high resolution images 
(compared to IRS~16NE and 16NW). 
Errors are around 30\% for sources of more than 0.5 Jy and 50\% for the sources with
flux densities below that value. The H-, K- and L-band values have errors up to 75\%. 
For the calculations, zero flux densities from Skinner et al. (2997) were used 
(FD$_{0,H}$ = 1020 Jy, FD$_{0,K}$ = 657 Jy, FD$_{0,L}$ = 253 Jy). The colors H-K and K-L are shown.}
\scriptsize
\label{table:3}      
\centering          
\begin{tabular}{l| c c c| c c  | c c c c c c c c c }     % 7 columns 
\hline\hline       

\scriptsize
Name & m$_{H}$ & m$_{K}$ & m$_{L}$ & H-K & K-L & FD$_{1.6\mu m}$ & FD$_{2.1\mu m}$ & FD$_{3.8\mu m}$ & FD$_{4.7\mu m}$ & FD$_{8.6\mu m}$ & FD$_{11.3\mu m}$ & FD$_{12.8\mu m}$ & FD$_{18.7\mu m}$ &FD$_{19.5\mu m}$ \\    % table heading 
\hline                        % inserts single horizontal line
IRS~5	   &	-- 		& -- 			& 4.4 	& -- & -- &  -- 		& -- 		& 4.270 & 4.880  & 5.110  & 4.210 & 5.850 & 3.210 & 2.850 \\
IRS~5NE  &	14.8 	& 12.7 		& 9.5   & 2.1 & 3.2 & 	0.001  & 0.006  & 0.047 & 0.800  & 0.510  & 0.600 & 0.560 & 0.950 & 0.740 \\
IRS~5E	 &	14.6	& 13.0 		& 10.7  & 1.6 & 2.3 &	0.002  & 0.004  & 0.016 & 0.210  & 0.590  & 1.060 & 1.540 & 1.580 & 1.720 \\
IRS~5S	 &	14.3  & 12.4 		& 9.9   & 1.8 & 2.5 &	0.002  & 0.007  & 0.032 & 0.430  & 0.630  & 0.510 & 0.380 & 0.630 & 1.060 \\
IRS~5SE1 &	-     & -	  	& 12.1	& --  & --  &	--     & --     & 0.004 & 0.180  & 0.500  & 0.970 & 1.100 & 2.840 & 3.580 \\
IRS~5SE2 &	13.2  & 11.6            & 10.5 	& 1.7 & 1.0 &	0.005  & 0.016 & 0.018 		& --     & 0.040  & 0.090 & 0.130 & --    & --    \\ \hline 
IRS~13N &	13.0 & 12.4   		& 10.0 	& 0.6 & 2.4 &	0.006  & 0.007 & 0.028 	& 1.580    & 0.380  & 0.650 	& 1.110 & 0.550    & 0.580    \\ \hline \hline
\end{tabular}
\end{table*}

\normalsize
   
To derive the stellar types of the observed sources we compared the reduced spectra shown in Fig.~\ref{FigVibStab} with standard star templates from \cite{Wallace} of spectral types from O- to M- and different luminosity classes. Since the template spectra are continuum normalized to one, we had to multiply them by a blackbody of temperature equal to the effective temperature of each stellar type. The spectra were also convolved with a Gaussian distribution to match the resolution of our observed spectra.
\\
The comparison was done by fitting the continuum of the source 
spectra (from 2.0 to 2.29~$\mu$m) with A$_{K}$ extincted template 
spectra, the apparent magnitude m$_{K}$, the depth $d$ and the equivalent 
width (EW) of the $^{12}$CO absorption bandheads. 
The fitting procedure is explained hereafter.
\\
A small amount of apparent residual broadening of the CO bandheads may be 
due to the following fact: 
Most of the sources are
apparently associated with small clusters of fainter stars.
As the spectra are taken against the underlying stellar 
cluster in moderate seeing in a 0.6'' diameter slit 4 to 6 sources that are 
individually up to 2 magnitudes
fainter than the program source may lead to a systematic broadening of the order of the 
cluster velocity dispersion of 1$\sigma$$\sim$150 km/s (corrseponding to a 3$\sigma$ value of
up to $\lambda$/$\Delta$$\lambda$$\sim$670$\sim$3-4 nm at this position. 
In addition, the spectra are of lower SNR than those of the template stars.
Since our fitting procedure is not solely based on the CO bandhead absorption, 
but also the extinction and apparent brightness of the objects, 
we consider the influence of these contamination effects to be 
negligible within the required accuracy of the procedure. 

\textbf{1.} The visual extinction in the line of sight to the GC varies 
between 20 and 50~mag with an average of around 30~mag 
(\cite{Rieke}; see also the extinction map presented by \cite{Schoedel2007} and \cite{Scoville}). It is mostly due to material along the line of sight to the GC, located, partly, in the the diffuse ISM (\cite{Lebofsky79}) and partly, in the dense molecular 
clouds (\cite{Gerakines}). In the K-band, the average extinction is A$_{K}$ = 0.112$\cdot$A$_{V}\simeq$ 3.4 mag 
(\cite{Rieke2}; \cite{Krabbe}; \cite{Scoville}).\\ 
In order to fit the observed spectra with those of template stars, we had to consider the foreground extinction. Therefore, we followed the same procedure as in Moultaka et al. (2004) and applied the extinction law of Martin et al. (1990) to redden the template spectra. An extincted spectrum is thus obtained as follows:
\begin{center}
f$_{ext,\lambda}$ = f$_{0,\lambda}\cdot$10$^{-0.4 \cdot A_{K} \cdot (2.2/\lambda)^{1.8}}$
\end{center}
with f$_{ext,\lambda}$ and f$_{0,\lambda}$ as the extinct and intrinsic flux densities, respectively, and $\lambda$ the wavelength in $\mu$m. The fitting process was done in the wavelength range from 2.07 to 2.29~$\mu$m, excluding the locations of detected prominent 
line features such as Br$\gamma$. We used this spectral range because it is mainly dominated by quasi-linear blackbody continuum and harbors no or few absorption and emission lines.\\ 
As a result of the unclear distribution of the K-band absorption due to the mini-spiral (\cite{Schoedel2007}), the fitting was done 
using K-band extinctions ranging from 2.4 to 4.4 magnitudes.
\\

\textbf{2.} The apparent K-band magnitudes of the standard stars at the distance of the GC (r = 8000 pc; \cite{Eisenhauer03}) are obtained via
\begin{center}
m$_{K}$ = M$_{K}$ + 5$\cdot$log~r - 5 $\approx$ M$_{K}$ + 14.52
\end{center}
where M$_{K}$ = M$_{V}$ - (V-K). The absolute magnitudes and colors are taken from Lang (1992) and Allen (2000).\\
The apparent magnitudes of the template stars are compared to the extinction corrected K-band magnitudes of the observed sources shown in Tables~\ref{table:3} and \ref{table:A}.\\

\textbf{3.} All sources clearly show the rotational vibrational $^{12}$CO(2-0) absorption 
bandhead at 2.2935~$\mu$m. It is the strongest absorption feature in K-band and 
depends, among other parameters, on the temperature of the objects 
and is therefore a good tracer of spectral types.\\
Gaffney et al. (1993; 1995)
show that the shape of this feature is fully characterized by its EW. 
Therefore, in the fitting process, the depth $d$ and the EW of this absorption feature 
are also considered as parameters to be fitted. 
The CO 2-0 band head depth is proportional to 1/T only within a luminosity class
(e.g. \cite{Orig}; \cite{Kleinmann}).
Based on the temperature and luminosity range (see Table~\ref{table:A}) we assume 
that the sources mostly belong to the luminosity class III 
and are therefore more likely giants than super giants or a mixture of classes.

The calculations were done via:
\begin{center}
$d$ = $\Delta\lambda\cdot$(1-f$_{min}$),\\
EW = $\int_{\lambda_{1}}^{\lambda_{2}}$(1-$f_{\lambda}$)d$\lambda$=$\Delta \lambda \sum_{\lambda_{1}}^{\lambda_{2}}$(1-f$_{\lambda}$)
\end{center}
where f$_{\lambda}$ is the normalized flux density in the wavelength range from $\lambda_{1}$=2.2924 to $\lambda_{2}$=2.2977~$\mu$m (\cite{Orig}) and $\Delta\lambda$ the wavelength bin size (=0.72 nm). f$_{min}$ is the normalized flux density of the deepest absorption in the described wavelength range.\\ 

\begin{table}
\caption{Measured parameters of the observed and template spectra (taken from \cite{Wallace}) and physical parameters of the template stars.}             % title of Table
\label{table:A}      % is used to refer this table in the text
\centering                          % used for centering table
\begin{tabular}{l |c c c |l c}         % centered columns (4 columns)
\hline\hline                 % inserts double horizontal lines
Object	 & EW 	& d			&	m$_{K}$		& Stellar & Temp.\\
	 &  [nm]	& 	[nm]		&	[mag]		& Type &  [K] \\ \hline
IRS~5NE	 & 0.239			& 0.892			& 13.2						& -- 	 & --					\\		
IRS~5E	 & 0.365 		  & 0.849			& 13.6						& -- 	 & --					\\				
IRS~5S	 & 0.326			& 0.885			& 12.9						& -- 	 & --					\\			
IRS~5SE	 & 0.188			& 0.931			& 11.8						& -- 	 & --					\\ \hline		
HR 2456  & 0.099			& 0.974			& 10.3						& O7V   & 34600	   \\
HR 6165  & 0.028			& 0.966			& 11.4						& B0V	  & 28000	   \\
HR 5191  & 0.059			& 1.000			& 13.5						& B3V   & 22570	\\
HR 3323  & 0.146		  & 0.904			& 13.3						& G5III & 4400	\\
HR 3212  & 0.256		  & 0.848			& 13.2						& G7III & 4286	\\
HR 2985  & 0.283		  & 0.834			& 13.2						& G8III & 4229	\\
HR 6703  & 0.226			& 0.846			& 13.2						& G8III & 4229	\\
HR 8317  & 0.516			& 0.755			& 12.9						& K0III & 4114	\\
HR 8694  & 0.407			& 0.773			& 12.9						& K0III & 4114	\\ \hline \hline
\end{tabular}
\end{table}

\begin{table}
\caption{The stellar templates that best fit the observed spectra. Listed are the K-band extinction needed in the best fit and the differences of the fitted parameters between the templates and the observed sources. All but IRS~5SE can be represented by a G-type giant. See text for further comments.}             % title of Table
\label{table:B}      % is used to refer this table in the text
\centering                          % used for centering table
\begin{tabular}{l| l c |c c c c}        % centered columns (4 columns)
\hline\hline                 % inserts double horizontal lines
Object		&	Stellar				&	Temp.					& $\Delta$EW 	& $\Delta$d & $\Delta$m$_{K}$ & A$_{K}$ \\
					&	Type		&	[K]						&	[nm] 				&	[nm]			& [mag] 					& [mag] \\ 	
\hline                        % inserts single horizontal line
IRS~5NE   &	G8III						&	4229					&	0.013 			& 0.046			& 0.0							& 3.4 		\\
IRS~5E		&	G8III						&	4229					&	0.082				& 0.015			& 0.4     				& 3.3 	  \\
IRS~5S		&	G7III						&	4286					&	0.070 			& 0.037     & 0.3							& 3.3 	\\
IRS~5SE	  &	B3V							&	22570					&	0.129 			& 0.069			& 1.7 						& 2.7 	\\ 
\hline \hline                                  %inserts single line
\end{tabular}
\end{table}

We selected the extincted templates (\cite{Wallace}) with the smallest difference in EW, bandhead depth $d$ 
and apparent magnitude m$_{K}$ compared to the observed spectra (see Table~\ref{table:A}). 
The stellar template spectra that fit the observed ones best are shown in Fig.~\ref{FigVibStab2}. 
Their physical parameters as well as the value of the K-band extinction A$_{K}$ needed for the best fit 
are listed in Table~\ref{table:B}.\\
All observed objects except IRS~5SE can be represented by late-type 
giants with surface temperatures in the range of 
log(T) = 3.6-3.7 (or 4000~K $<$ T $<$ 5000~K)
and with apparent K-band magnitudes of 10 to 14~mag. These values are consistent with the expected properties of stars on the AGB at the distance of the GC assuming an average K-band extinction of 3.4~mag (see also Fig.~14 in \cite{Rafelski}). The only exception is IRS~5SE. From our fits we derive that for IRS~5SE the temperature is 
likely to be higher than 5000~K, but in this case the template stars miss to fit the observed apparent 
brightness by about 2 magnitudes. This is a strong indication that we observe in fact a blend of an early- and a late-type star and that the observed CO bandheads are correspondingly too weak compared to the continuum - resulting in the described
discrepancies.

\subsection{Spectral energy distributions, colors}

\begin{figure*}
\centering
\includegraphics[width=18cm]{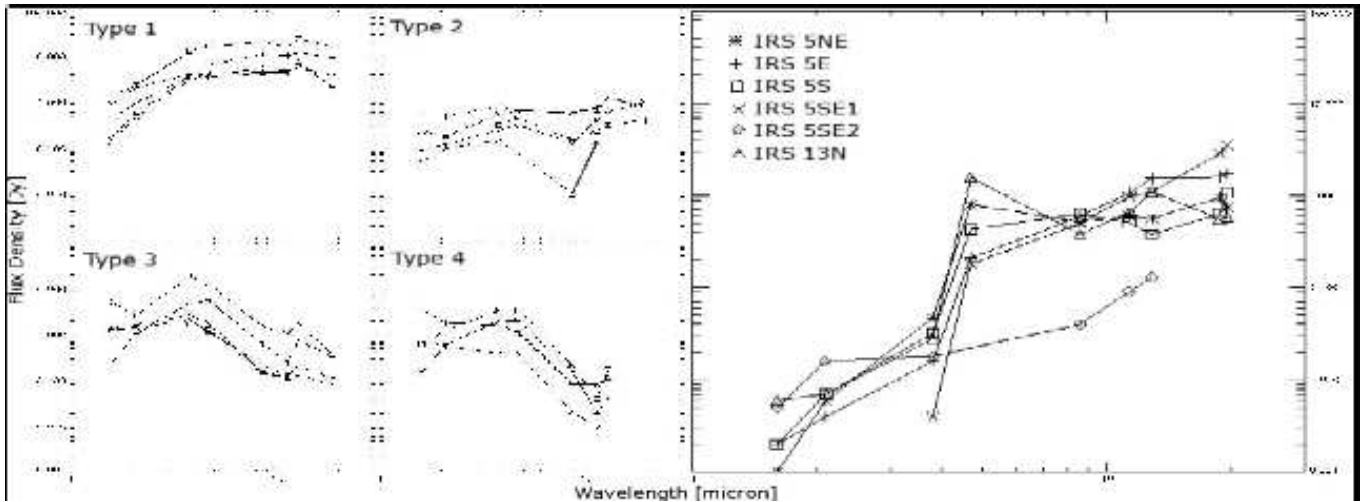}
\caption{The SEDs of the different types of MIR-sources introduced by Viehmann et al. (2006) 
compared to the five 
compact MIR sources located east of IRS~5 and the unclassified source IRS~13N:
Type~1: luminous northern arm bow shock sources, 
Type~2: lower luminosity bow shock sources, 
Type~3: cool stars 
and 
Type~4: hot stars.
For the SEDs shown in the right panel see data and error estimates in Table~\ref{table:3}      
and the corresponding caption.
There is a strong onset of dust emission longward of 2.2~$\mu$m in these sources.
The strongest change is present in the IRS13N sources (see Eckart et al. 2004).
For sources without such a strong IR-excess (e.g. IRS 5 SE2)
the transition from the non-thermal to the thermal IR is smoother.
}
         \label{FigVibStab5}
   \end{figure*}

\begin{figure}
   \centering
   \includegraphics[width=7cm]{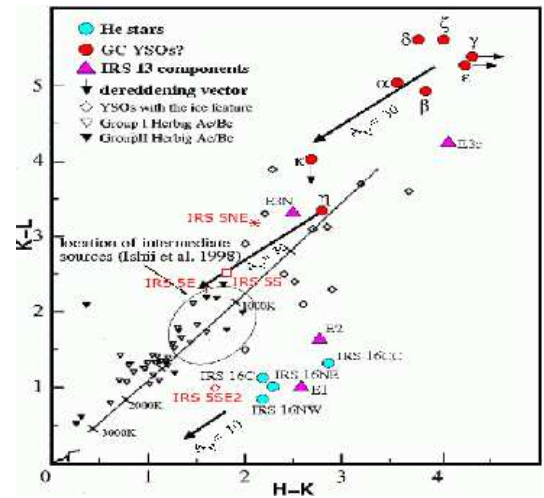}
      \caption{HKL two-color diagram (\cite{Eckart}) showing values of H-K and K-L of L-band excess sources in the IRS~13N cluster ($\alpha$, $\beta$, $\gamma$, $\delta$, $\epsilon$, $\zeta$, $\eta$ and $\kappa$), He-stars, young stellar objects (YSOs) with ice features, Herbig Ae/Be stars of group I and II and some intermediate sources observed by Ishii et al. (1998).
The locations of the sources studied in this paper are also shown. The dash-dotted line shows the colors of a single blackbody at different temperatures. An extinction vector is shown for a visual extinction of A$_{V}$ = 10~mag as well as A$_{V}$ = 30~mag.}
         \label{FigVibStab20}
   \end{figure}

In a recent paper (\cite{Viehmann06}), we described spectral energy distributions (SEDs) from H- to Q-band (1.6 to 19.5~$\mu$m) to investigate colors, infrared excesses and extended 
dust emission of GC sources. 
The SEDs showed characteristic features. 
We could clearly distinguish between four types of sources (Fig.~\ref{FigVibStab5}): 
the luminous northern arm bow shock sources (Type~1), the lower luminosity bow shock sources (Type~2), 
the cool stars (Type~3) and the hot stars (Type~4).
[Note: Although the lower Type~2 source has a stronger curved short wavelength 
spectrum - and therefore looks more like a Type~3 source in that spectral 
domain - it is clearly identified as a lower luminosity bow shock source 
(see Viehmann et al. 2006, Clenet et al. 2004).]
This classification may be used to clarify the nature of the presently unclassified sources 
east of IRS~5 discussed in this work. The H-, K- and L-band luminosities were derived from the comparison of IRS~13N and the sources east of IRS~5 with IRS~16NE and 16NW (m$_{K}$, H-K and K-L from \cite{Blum2}). For the calculations (m=-2.5$\cdot$log(FD$_{1,\lambda}$/FD$_{0,\lambda}$)), zero flux densities from \cite{Skinner} were used (FD$_{0,H}$ = 1020 Jy, FD$_{0,K}$ = 657 Jy, FD$_{0,L}$ = 253 Jy).
The compact mid-IR sources east of IRS~5
are less luminous and situated close to or within the northern arm of the mini-spiral and therefore 
show similar characteristics to identified lower luminosity bow shock sources. 
However, with the exception of IRS~5SE, their appearance on high resolution K- and L-band NAOS/CONICA 
images obtained with adaptive optics does not indicate bow shock structures (see Fig.~\ref{FigVibStab8} and e.g. 
\cite{Moultaka04}; \cite{Moultaka05}; \cite{Viehmann05}; \cite{Clenet}). Moreover, all our sources show SEDs that are either flat between 4.7 and 20~$\mu$m, like the Type~2 sources of Viehmann et al. (2006) (IRS~5NE and IRS~5S, see Fig.~\ref{FigVibStab5}), or increase towards longer wavelengths (IRS~5E, IRS~5SE1 and IRS~5SE2). This is similar to IRS~13N where a strong foreground extinction and/or dust emission (T $>$ 500 K) was assumed (\cite{Eckart}). The object could be identified as a cluster of stars that heat the local environment of the mini-spiral or young stars with ages of 0.1 to 1~Myr.\\
The sources east of IRS~5 show very weak luminosities in H- and K-band. 
In the NIR two-color diagram (Fig.~\ref{FigVibStab20}) IRS~5NE, IRS~5E and IRS~5S 
are located close to the positions of 
group II Herbig Ae/Be stars with ages of about 0.1 to 1 million yr
after correction for the 30~mag of the foreground extinction.
These Herbig Ae/Be stars are young stars of intermediate mass (2-8~M$_{\odot}$), 
embedded in dust that is not confined in disks (e.g. \cite{Hillen}). 
Since the sources east of IRS~5 show similar SEDs as the IRS~13N sources and their K-band spectra are fitted by intermediate mass giants, their colors and SEDs are most probably due to the dust they are embedded in. IRS~5SE1 is less luminous in the NIR and MIR. This matches the fit of the spectrum by a giant of higher temperature than the other sources east of IRS~5.

\subsection{Emission lines}
The resolution in our spectra is low.
Therefore we used simple Gaussian fits to derive
the emission line fluxes. We list the fluxes F and line widths $v$ (corrected for the spectral resolution)
of the detected features in Table~\ref{table:5}.
In this table, the uncertainties take into account that the Pa$\alpha$ line is located in a spectral region of lower SNR. 
In general, no emission
lines are expected, especially if the stellar source is of late type. 
Therefore, a different source in the line-of-sight must be the origin of the emission lines. 
The gas and dust of the circum-stellar media and/or the northern arm of the mini-spiral could be responsible for all detected emission lines: the consistent average line widths suggest the same source for all emission lines.\\

\begin{table}
\caption{Fluxes and line widths (and their 1$\sigma$ uncertainties) 
of the detected emission lines for the different observed K-band sources.
The line fluxes of Br$\delta$ and Pa$\alpha$ suffer from 
atmospheric calibration problems and have to be taken with caution.
For spectral flux calibration - see section~2.2.
}
\label{table:5}      
\centering          
\begin{tabular}{l | l c c}     % 7 columns 
\hline\hline       
Object & Line 	      	& F$\pm \Delta$F 	& v$\pm \Delta$v   \\
     & Feature      	& [10$^{-19}$W/m$^{2}$] 	& [km/s] \\                 % table heading 
\hline                        									% inserts single horizontal line
IRS~5NE  & Pa $\alpha$		& 551$\pm$93 	& 400$\pm$384	\\
	& Br $\delta$ 		& 50$\pm$8 	& 554$\pm$370	\\
	& Br $\gamma$ 		& 50$\pm$7 	& 356$\pm$332	\\ \hline
IRS~5E	 & Pa $\alpha$		& 649$\pm$147 	& 390$\pm$384	\\
	& Br $\delta$ 		& 47$\pm$9 	& 492$\pm$370	\\
	& Br $\gamma$		& 87$\pm$18 	& 357$\pm$332	\\ \hline
IRS~5S	 & Pa $\alpha$		& 649$\pm$106 	& 490$\pm$384	\\
	& Br $\delta$ 		& 45$\pm$7 	& 493$\pm$370	\\
	& HeI			& 21$\pm$3 	& 428$\pm$349	\\
	& Br $\gamma$		& 72$\pm$11 	& 402$\pm$332	\\ \hline
IRS~5SE  & Pa $\alpha$		& 770$\pm$176 	& 489$\pm$384	\\
	& Br $\delta$		& 96$\pm$15 	& 508$\pm$370	\\
	& HeI			& 29$\pm$4 	& 532$\pm$349	\\
	& Br $\gamma$		& 98$\pm$15 	& 429$\pm$332	\\
\hline  \hline                  
\end{tabular}
\end{table}

%\textbf{1.} {\it Hydrogen and Helium emission lines:}
The hydrogen recombination lines Pa$\alpha$ (1.8756~$\mu$m) and Br$\gamma$ (2.1661~$\mu$m) are the most prominent emission features in all observed spectra. Like the HeI line at 2.0587~$\mu$m, they either arise in gas in the vicinity of hot early type stars or in the more extended gas distribution of the mini-spiral.
In this case the gas is ionized and heated by the UV radiation field in the central parsec
and therefore also linked to the presence of the massive 
and hot He-stars. 

%\textbf{2.} {\it Molecular hydrogen emission lines:}
%, but not the H$_{2}$ 1-0 S(1) transition at 2.1213~$\mu$m. Since the latter is not detected in the spectra, we could only estimate an upper limit of the line flux ratio F$_{H_{2}1-0S(1)}$/F$_{Br \gamma}$ of less than 0.2. This indicate that both lines are expected to be partly excited by the UV light of the inner hot OB stars or by collisional shocks.

\section{Discussion}
Here we discuss the properties of the compact MIR sources east of IRS~5 based on 
their spectra, proper motions as well as their clustering and possible interaction with the 
local ISM of the GC.

\textbf {Proper motions:}
The compact MIR excess sources close to the northern arm of the mini-spiral 
show proper motions to the east-southeast and northwest.
These motions are different from the overall motion of the gas contained in the northern arm
which moves predominantly  southward in this section of the mini-spiral
(\cite{Paumard06}; \cite{Muzic}).
However, the motions of the compact sources may well be consistent with the motion of objects that are 
located in the clockwise and counter clockwise rotating stellar disks of the central cluster.
In that case, the spatial extent of the two stellar disks would be larger than currently assumed
(\cite{Genzel03}; \cite{Paumard06}).
The IRS~5E and IRS~5NE sources show proper motions towards the northwest.
This is in agreement with the overall motion of sources 
in the clockwise rotating disk of stars. 
%In the IRS~5 area we find here objects moving to the west and north west. 
If the compact MIR bright objects east of IRS~5 were in interaction with the mini-spiral 
they would also be located behind the center of the stellar cluster - as the mini-spiral is
in this region. 
IRS~5S1 to 3 and IRS~5SE1 show proper motions to the east-southeast. 
This is in agreement with the overall orientation of the
counter clockwise rotating disk of stars. 
%Here we find in the general IRS~5 region objects moving to the west or southwest. 
This stellar disk also has a higher velocity dispersion compared to the 
clockwise rotating disk, such that the sources could also be located 
behind the center of the stellar cluster.
Given the width of the mini-spiral and the apparent separation and proper motion of 
the sources their possible interaction with the mini-spiral must still be ongoing 
or must have been very recent (less than a few hundred years ago).\\

A comparison to more luminous and well studied sources in the GC stellar cluster 
suggests four plausible explanations for the sources discussed here:

\textbf{1.} They could be young stellar objects (YSOs) which have been formed while falling 
into the central parts of the 0.3 pc core radius stellar cluster. These objects 
could be bright in the MIR since they are still surrounded by their proto-stellar
dust shells or disks which are about to be lost in the dense stellar environment 
of the central cluster. 
If we find evidence for this hypothesis it will have a profound influence 
on the yet unsolved formation process of young stars in the central cluster.

\textbf{2.} They could be low luminosity counterparts of the more luminous bow shock sources 
(i.e. IRS~5, 10W) within the central 5$''$. 

\textbf{3.} They could be a more dispersed group of stars similar to the IRS~13N complex 
described by \cite{Eckart} (Fig.~\ref{FigVibStab8}). They are more luminous than the 
sources in the IRS~13N complex which may be due to the fact that they are less extinct,
but show similar SEDs and colors.

\textbf{4.} They could be lower luminosity but dust forming AGB stars.\\

\textbf {IRS~5NE,~5E,~5S are difficult to explain as AGB stars:}
Several bright (m$_{K}$ = 10 to 12 mag) AGB stars represent both an 
intermediate-mass and an intermediate-age component of the GC stellar cluster with an age of about 100 Myr 
(\cite{Lebofsky87}; \cite{Krabbe}; \cite{Blum}; \cite{Genzel03}). 
In addition to bright blue super giants (in the IRS~16 and IRS~13 complexes) and
red super giants (IRS~7) these AGB stars dominate the H- and K-band images.
Prominent representatives are IRS~12N, IRS~10EE and IRS~15NE.\\
A possible scenario is that the observed late type sources that are MIR bright and associated with
dust emission are AGB stars up to 3 magnitudes fainter than their brighter
representatives in the central stellar cluster (see above).
Luminous and dust-enshrouded 
AGB stars can be found close to the tip of the AGB. 
They will not evolve significantly in luminosity before mass loss ends their AGB evolution. 
The total span of mass-loss rates which can be derived from observations 
reaches from 10$^{-7}$ to 10$^{-3}$ \solm/yr (\cite{vanLoon})
with typical wind velocities of the order of 
10-20 km/s (e.g. \cite{Bergeat}; \cite{Hagen}).
In general, more luminous and cooler stars are found to reach higher mass-loss rates. 
Stellar model calculations (\cite{Schroeder}) show 
a collective mass-loss rate of 5.0$\times$10$^{-4}$ \solm/yr 
for a synthetic sample of more than 5000 brighter tip-AGB stars, 
which makes them excellent candidates for MIR excess emission. 
Also Tanner et al. (2002; see their Table~4) have pointed out that such AGB stars
could be bow shock sources with winds of up to 40~km/s and appreciable stand off distances.
Therefore IRS~5NE, 5E and 5S may be AGB stars and a significant amount of the MIR
excess may be due to dust shells produced by the individual sources.\\

\noindent
There are, however, indications that the sources are young:

\textbf {Indications for a bow shock source near IRS~5SE:}
Our measurements allowed us to narrow down the stellar types of each observed source neglecting 
the influence of the mini-spiral. 
With the exception of IRS~5SE all sources show characteristics of K-type giants with T $<$ 5000 K. 
Here, however, the reduced K-band spectrum is a composition of the spectra of the 
individual sources IRS~SE2 and IRS~SE1. IRS~SE2 is blue and brighter than IRS~SE1.
Following the analysis of the overall spectrum of IRS~5SE in section~\ref{Stellartypes}
we therefore identify IRS~5SE2 as a hot early type star of at least spectral type B3
and attribute the CO bandhead contained in the overall spectrum of IRS~SE 
to the fainter component IRS~SE1.
Due to confusion with IRS5~SE2 the CO bandhead of the western source IRS5~SE1, 
a cool late type star that is bright in N- and Q-bands, is diluted and 
therefore the stellar template we derive suggests an earlier type source. 
All observed sources show UV excited emission features from the gas and dust clouds 
they are embedded in. In all of the cases we cannot exclude that the dominant part of 
the hydrogen recombination lines are in fact due to emission from the mini-spiral in that region.
Since the spectrum of IRS~5SE shows contributions from an early type star, and a 
bow shock structure is visible, superimposed on IRS~5E1, we conclude that IRS~5SE should be classified
as a lower luminosity bow shock source.
\\

\textbf {Clustering might indicate young sources:}
Based on their distribution and proper motion the
sources associated with IRS~5NE and IRS~5SE as well as the two bright IRS~5E objects
are not likely to be clustered. 
However, the IRS~5S1-3 objects and possibly the faint objects associated 
with IRS~5E1  show an indication for a clustering of the sources. 
Fig.~\ref{figpm1} shows that there are at least 5 stars brighter than about K=15
which are located within about 250~mas of IRS~5S. Three of them show within the uncertainties, a common 
motion towards the east-southeast (+120~km/s in R.A.
and -30~km/s in Dec.; see Table~\ref{pmtab}). 
This suggests that they may in fact be physically associated.
Clustering of sources is apparently a frequent phenomenon in the field east of IRS~5 and in the 
GC stellar cluster in general 
(e.g. \cite{Schoedel2007}; \cite{Schoedel2005}; \cite{Eckart}; \cite{Genzel03}).
As can be seen from Table~\ref{pmtab} and Figs.~\ref{figpm1} and  \ref{figpm2} there are 
at least two more candidates for source clustering in the field ($\#$14a-e and $\#$21a-c).
However, these source complexes are not as prominent in the MIR as the four sources east of 
IRS~5 described in this paper.
Clustering of sources is usually taken as an indication of systems that are young (at least dynamically).\\

If the clustering of the sources and their possible association with
the clockwise and counter-clockwise stellar disks is taken as an indication of youth, 
this is in conflict with the interpretation of IRS~5NE, IRS~5E and IRS~5S as only moderately bright
AGB stars, and with their observed spectra. 
The location of the sources in the two-color-diagram at the position of 
Herbig Ae/Be group II stars also supports that the sources are young, with ages below one million years.
In addition, their interaction with the
mini-spiral cannot be excluded and could be a reason for their MIR brightness.
This would also be fully consistent with their identification as low luminosity bow shocks
from NIR/MIR SEDs. 
In this case, however, the CO bandhaeads in the spectra are difficult to explain.

\section{Summary}
We have presented new imaging, proper motion, and spectroscopic data on a class of 
compact MIR sources in the GC.
The brightest sources contained in IRS~5NE, 5E and 5S may be 
AGB stars and a part of the MIR
excess may be due to dust shells produced by the individual sources. 
They are, however, fainter than expected for AGB stars in the GC stellar cluster. 
In all cases an interaction with the mini-spiral cannot be excluded and their broad band
infrared SEDs indicate that they could be
lower luminosity counterparts of the identified bow shock sources,
and at least in the case of IRS~5S and IRS~5E1 -
may belong to dynamically young associations of stars.
The blue star IRS~5SE2 is that best candidate associated with a bow shock source.

\begin{acknowledgements}
      Part of this work was supported by the German
      \emph{Deut\-sche For\-schungs\-ge\-mein\-schaft, DFG\/} via grant SFB 494.
K. Muzic is member of the International Max Planck Research School (IMPRS) for 
Radio and Infrared Astronomy at the MPIfR and the Universities of Bonn and Cologne.
\end{acknowledgements}


\begin{thebibliography}{}

  \bibitem[Allen]{Allen} Allen, C.W., Allen's Astrophysical Quantities, Cox, Arthur N. (Ed.), Athlone Press, London, UK, 1976 4th ed. 2000.
  \bibitem[Becklin \& Neugebauer 1968]{Becklin1} Becklin, E. E.; Neugebauer, G. 1968, ApJ, 151, 145 
  \bibitem[1969]{Becklin2} Becklin, E. E.; Neugebauer, G. 1969, ApJ, 157, L31
  \bibitem[Bergeat \& Chevallier 2005]{Bergeat} Bergeat, J.; Chevallier, L. 2005, A\&A 429, 235-246 
  \bibitem[Blum et al. 1996]{Blum} Blum, R. D.; Sellgren, K.; Depoy, D. L 1996, AJ, 112, 1988
  \bibitem[Blum et al. 1996]{Blum2} Blum, R. D.; Sellgren, K.; Depoy, D. L. 1996, ApJ, 470, 864
  \bibitem[Brandner et al. 2002]{brandner02} Brandner, W.; Rousset, G.; Lenzen, R.; Hubin, N.; Lacombe, F.; Hofmann, R.; Moorwood, A. 2002, ESO Messenger, Vol.107, p.1
  \bibitem[Carr et al. 2000]{Carr} Carr, J. S.; Sellgren, K.; Balachandran, S. C. 2000, ApJ, 530, 307
  \bibitem[Clenet et al. 2004]{Clenet} Clenet, Y.; Rouan, D.; Gratadour, D.; Lacombe, F.; Gendron, E.; Genzel, R.; Ott, T.; Sch\"odel, R.; Lena, P. 2004, A\&A, 424, L21
  \bibitem[Eckart et al. 1990]{eckart90} Eckart, A.; Duhoux, P. R. M. 1990, ASPC, 14, 336 
  \bibitem[Eckart et al. 2004]{Eckart} Eckart, A.; Moultaka, J.; Viehmann, T.; Straubmeier, C.; Mouawad, N. 2004, ApJ 602, 760
  \bibitem[Eisenhauer et al. 2003]{Eisenhauer03} Eisenhauer, F.; Sch\"odel, R.; Genzel, R.; Ott, T.; Tecza, M.; Abuter, R.; Eckart, A.; Alexander, T. 2003, ApJ, 597, 121
  \bibitem[Eisenhauer et al. 2005]{Eisenhauer05} Eisenhauer, F.; Genzel, R.; Alexander, T.; Abuter, R.; Paumard, T.; Ott, T.; Gilbert, A.; Gillessen, S.; Horrobin, M.; Trippe, S.; and 11 coauthors 2005, ApJ, 628, 246
  \bibitem[Gaffney \& Lester 1993]{Gaffney1} Gaffney, N. I.; Lester, D. F.; Telesco, C. M. 1993, ApJ, 407, L57
  \bibitem[1995]{Gaffney2} Gaffney, N. I.; Lester, D. F.; Doppmann, G. 1995, PASP, 107, 68
  \bibitem[Geballe et al. 2004]{Geballe1} Geballe, T. R.; Rigaut, F.; Roy, J.-R.; Draine, B. T. 2004, ApJ, 602, 770
  \bibitem[Geballe et al. 2006]{Geballe2} Geballe, T. R.; Najarro, F., Rigaut, F. et al., 2006, ApJ, 652, 370
  \bibitem[Genzel et al. 2003]{Genzel03} Genzel, R.; Sch\"odel, R.; Ott, T.; Eisenhauer, F.; Hofmann, R.; Lehnert, M.; Eckart, A.; Alexander, T.; Sternberg, A.; Lenzen, R.; and 5 coauthors 2003, ApJ, 594, 812 
\bibitem[Ghez et al.(2005)]{Ghez05} Ghez, A.M., Salim, S., Hornstein, S. D., Tanner, A., Lu, J. R., Morris, M., Becklin, E. E., Duch\^ene, G.,2005, ApJ 620, 744
  \bibitem[Gerakines et al. 1999]{Gerakines} Gerakines, P. A.; Whittet, D. C. B.; Ehrenfreund, P.; Boogert, A. C. A.; Tielens, A. G. G. M.; Schutte, W. A.; Chiar, J. E.; van Dishoeck, E. F.; Prusti, T.; Helmich, F. P.; de Graauw, Th. 1999, ApJ, 522, 357
  \bibitem[Hagen 1978]{Hagen} Hagen, W. 1978, ApJS, 38, 1
%  \bibitem[Hummer \& Storey 1987]{Hummer} Hummer, D. G.; Storey, P. J. 1987, MNRAS, 224, 801
  \bibitem[Hillenbrand et al. 1992]{Hillen} Hillenbrand, L. A.; Strom, S. E.; Vrba, F. J.; Keene, J. 1992, ApJ, 397, 613
  \bibitem[Ishii et al. 1998]{Ishii} Ishii, M.; Nagata, T.; Sato, S.; Watanabe, M.; Yao, Y.; Jones, T. J. 1998, AJ, 116, 868
%  \bibitem[Johnson 1966]{Johnson} Johnson, H. L. 1966, ARA\&A, 4, 193
  \bibitem[Kleinmann \& Hall 1986]{Kleinmann} Kleinmann, S. G.; Hall, D. N. B. 1986, ApJS, 62, 501
  \bibitem[Krabbe et al. 1995]{Krabbe} Krabbe, A.; Genzel, R.; Eckart, A.; Najarro, F.; Lutz, D.; Cameron, M.; Kroker, H.; Tacconi-Garman, L. E.; Thatte, N.; Weitzel, L.; and 4 coauthors 1995, ApJ, 447, L95
  \bibitem[Lang 1992]{Lang} Lang, K. R. 1992, Astrophysical Data I. Planets and Stars, X, Springer-Verlag Berlin Heidelberg New York
  \bibitem[Lebofsky 1979]{Lebofsky79} Lebofsky, M. J. 1979, AJ, 84, 324
  \bibitem[Lebofsky \& Rieke 1987]{Lebofsky87} Lebofsky, M. J.; Rieke, G. H. 1987, AIPC, 155, 79
  \bibitem[Lenzen et al. 1998]{lenzen98} Lenzen, R.; Hofmann, R.; Bizenberger, P.; Tusche, A. 1998, SPIE, 3354, 606 
  \bibitem[Lenzen et al. 2003]{lenzen03} Lenzen, R.; Hartung, M.; Brandner, W.; Finger, G.; Hubin, N. N.; Lacombe, F.; Lagrange, A.-M.; Lehnert, M. D.; Moorwood, A. F. M.; Mouillet , D. 2003, SPIE, 4841, 944
  \bibitem[Low et al. 1969]{Low} Low, F. J.; Kleinmann, D. E.; Forbes, F. F.; Aumann, H. H. 1969, ApJ, 157, L97 
  \bibitem[Martin \& Whittet 1990]{Martin90} Martin, P. G.; Whittet, D. C. B. 1990, ApJ, 357, 113
  \bibitem[Menten et al. 1997]{menten97} Menten, K.M.; Reid, M.J.; Eckart, A.; Genzel, R. 1997, ApJ, 475, L111
  \bibitem[Moorwood et al. 1998]{Moorwood} Moorwood, A.; Cuby, J.-G. 1998, Msngr, 94, 7 
  \bibitem[Moultaka et al. 2004]{Moultaka04} Moultaka, J.; Eckart, A.; Viehmann, T.; Mouawad, N.; Straubmeier, C.; Ott, T.; Sch\"odel, R. 2004, A\&A, 425, 529	
 \bibitem[2005]{Moultaka05} Moultaka, J.; Eckart, A.; Sch\"odel, R.; Viehmann, T.; Najarro, F. 2005, A\&A, 443, 163
 \bibitem[Muzic et al. 2007]{Muzic} Muzic, K.; Eckart, A.; Schoedel, R.; Meyer, L.; Zensus, A., 2007, A\&A 469, 993
  \bibitem[Najarro et al. 1997]{Najarro} Najarro, F.; Krabbe, A.; Genzel, R.; Lutz, D.; Kudritzki, R. P.; Hillier, D. J. 1997, A\&A, 325, 700
  \bibitem[Origlia, Moorwood \& Oliva 1993]{Orig} Origlia, L.; Moorwood, A. F. M.; Oliva, E. 1993, A\&A, 280, 536
  \bibitem[Osterbrock et al. 2006]{Osterbrock} Osterbrock, D. E.; Ferland, G. J. 2006, 
      Astrophysics of gaseous nebulae and active galactic nuclei, 2nd. ed. by D.E. Osterbrock and G.J. Ferland.,
      Sausalito, CA: University Science Books, 2006
  \bibitem[Ott et al. 1999]{Ott} Ott, T.; Eckart, A.; Genzel, R. 1999, ApJ, 523, 248
  \bibitem[Paumard et al. 2006]{Paumard06} Paumard, T.; Genzel, R.; Martins, F.; Nayakshin, S.; Beloborodov, A. M.; Levin, Y.; Trippe, S.; Eisenhauer, F.; Ott, T.; Gillessen, S.; and 4 coauthors 2006, ApJ, 643, 1011	
  \bibitem[Rafelski et al. 2007]{Rafelski} Rafelski, M.; Ghez, A.M.; Hornstein, S.D.; Lu, J.R.; Morris, M. 2007, ApJ, 659, 1241
  \bibitem[Reid et al. 2003]{reid03} Reid, M.J.; Menten, K.M.; Genzel, R.; Ott, T.; Sch\"odel, R.; Eckart, A. 2003, ApJ, 587, 208
  \bibitem[Rieke \& Lebofsky 1985]{Rieke2} Rieke, G. H.; Lebofsky, M. J. 1985, ApJ, 288, 618
  \bibitem[Rieke et al. 1989]{Rieke} Rieke, G. H.; Rieke, M. J.; Paul, A. E. 1989, ApJ, 336, 752
  \bibitem[Rigaut et al. 2003]{Rigaut} Rigaut, F.; Geballe, T. R.; Roy, J.-R.; Draine, B. T. 2003, ANS 324, 551
  \bibitem[Rousset et al. 1998]{rousset98} Rousset, G.; Lacombe, F.; Puget, P.; Hubin, N. N.; Gendron, E.; Conan, J.-M.; Kern, P. Y.; Madec, P.-Y.; Rabaud, D.; Mouillet, D.; and 2 coauthors 1998, SPIE, 3353, 508
  \bibitem[Rousset et al. 2003]{rousset03} Rousset, G.; Lacombe, F.; Puget, P.; Hubin, N. N.; Gendron, E.; Fusco, T.; Arsenault, R.; Charton, J.; Feautrier, P.; Gigan, P.; Kern, P. Y.; Lagrange, A.-M.; Madec, P.-Y.; Mouillet, D.; Rabaud, D.; Rabou, P.; Stadler, E. ; Zins, G., 2003, SPIE, 4839, 140
 \bibitem[]{}Sch\"odel, R., Ott, T., Genzel, R., Hofmann, R., Lehnert, M., Ecka
rt, A., Mouawad, N., Alexander, T., 2002, Natur 419, 694
  \bibitem[]{}Sch\"odel, R., Ott, T., Genzel, R., Eckart, A., Mouawad, N., Alexander, T., 2003, ApJ 596, 1015
  \bibitem[2005]{Schoedel2005} Sch\"odel, R.; Eckart, A.; Iserlohe, C.; Genzel, R.; Ott, T. 2005, ApJ, 625, L111
  \bibitem[Sch\"odel et al. 2007]{Schoedel2007} Sch\"odel, R.; Eckart, A.; Genzel, R.; Merritt, D.; Alexander, T.; Sternberg, A.; Moultaka, J.; Ott, T.; Straubmeier, C. 2007, astro.ph.03178
  \bibitem[Schr\"oder et al. 2003]{Schroeder} Schr\"oder, K.-P.; Wachter, A.; Winters, J. M. 2003, A\&A, 398, 229-237 
  \bibitem[Scoville et al. 2003]{Scoville} Scoville, N. Z.; Stolovy, S. R.; Rieke, M.; Christopher, M.; Yusef-Zadeh, F. 2003, ApJ, 594, 294
  \bibitem[Skinner 1997]{Skinner} Skinner, C. J. 1997, hstc.work, 233
  \bibitem[Tanner et al. 2002]{Tanner1} Tanner, A.; Ghez, A. M.; Morris, M.; Becklin, E. E.; Cotera, A.; Ressler, M.; Werner, M.; Wizinowich, P. 2002, ApJ, 575, 860
  \bibitem[2003]{Tanner2} Tanner, Angelle M.; Ghez, A. M.; Morris, M.; Becklin, E. E. 2003, ANS, 324, 597
  \bibitem[2005]{Tanner3} Tanner, A.; Ghez, A. M.; Morris, M. R.; Christou, J. C. 2005, ApJ, 624, 742
  \bibitem[van Loon et al. 1999]{vanLoon} van Loon, J. Th.; Groenewegen, M. A. T.; de Koter, A.; Trams, N. R.; Waters, L. B. F. M.; Zijlstra, A. A.; Whitelock, P. A.; Loup, C. 1999, A\&A, 351, 559-572  
  \bibitem[Viehmann et al. 2005]{Viehmann05} Viehmann, T.; Eckart, A.; Sch\"odel, R.; Moultaka, J.; Straubmeier, C.; Pott, J.-U. 2005, A\&A, 433, 117	
  \bibitem[Viehmann et al. 2006]{Viehmann06} Viehmann, T.; Eckart, A.; Sch\"odel, R.; Pott, J.-U.; Moultaka, J. 2006, ApJ, 642, 861
  \bibitem[Wallace \& Hinkle 1997]{Wallace} Wallace, L.; Hinkle, K. 1997, ApJS, 111, 445


\end{thebibliography}
\end{document}